\documentclass[epj]{svjour}
\usepackage{graphicx}
\usepackage{multirow}
\begin{document}
\title{Tensor force effects and high-momentum components
 in the nuclear symmetry energy}
\author{A. Carbone\inst{1} \and  A. Polls\inst{1} \and  C. Provid\^encia\inst{2} \and  A. Rios\inst{3} \and I. Vida\~na \inst{2} 
}                     
%
%
\institute{Department d'Estructura  i Constituents de la Mat\`eria and Institut de Ci\`encies del Cosmos, Facultat de F\'{\i}sica, Universitat de Barcelona, E-08028 Barcelona, Spain
\and Centro de F\'\i sica Computacional, Department of Physics, University of Coimbra, PT-3004-516 Coimbra, Portugal 
\and Department of Physics, Faculty of Engineering and Physical Sciences, University of Surrey, Guilford GU2 7XH, United Kingdom}
\date{Received: date / Revised version: date}
%
\abstract{We analyze microscopic many-body calculations of the nuclear symmetry energy and its density dependence. 
The calculations are performed  in the
 framework of the Brueckner--Hartree--Fock  and the Self--Consistent Green's Functions methods. 
Within Brueckner--Hartree--Fock, the Hellmann--Feynman theorem gives access to the kinetic energy contribution as well as the contributions of the different components of the nucleon-nucleon interaction. 
The tensor component  gives the largest contribution to the symmetry energy. 
The decomposition of the symmetry energy in a kinetic part and a potential energy
part provides physical insight on the correlated nature of the system, indicating that neutron matter is less correlated than symmetric nuclear matter. Within the Self--Consistent Green's Function approach, we compute the momentum distributions and we identify the effects of the high momentum components  in the symmetry energy. 
The results are obtained for  the realistic interaction Argonne V18 potential, supplemented by the Urbana IX three-body force in the Brueckner--Hartree--Fock calculations.
} 
\maketitle
%


\section{Introduction}
\label{intro}
The nuclear symmetry energy, defined as the difference between the energy per particle
 of pure neutron matter (PNM) and symmetric nuclear matter (SNM), 
and, in particular, its density dependence, is a crucial ingredient to understand many properties of isospin-rich 
nuclei and neutron stars \cite{chen2008,lattimer2005}.  
A major scientific experimental and theoretical effort is being devoted to study the properties of
 asymmetric nuclear systems. 
Laboratory experiments, such as those recently performed  or being
 planned in existing
or next-generation radioactive ion beam facilities such as 
the Facility for Antiproton and Ion Research (FAIR, Germany), 
Rikagaku Kenkyusho (RIKEN, Japan),  
the Heavy Ion Research Facility in Lanzhou (HIRFL, \linebreak China),
SPIRAL2 at the Grand Accelerateur National d'Ions Lourds (GANIL, France), 
and the upcoming Facility for Rare Isotope Beams (FRIB, Michigan State University) 
can probe the density behavior of the symmetry energy \cite{chen2008}.
More precisely, experimental information on the density dependence of the symmetry energy below, 
close to and above the saturation density of nuclear matter can be obtained from
isospin diffusion measurements \cite{chen2005}, giant \cite{garg2007} and
pigmy resonances \cite{klim2007}, isobaric analog states \cite{dani2009}, isoscaling \cite{shetty2007} or meson production in heavy ion collisions \cite{li2005,fuchs2006}. 
Moreover, the accurate measurements of the neutron skin thickness  in $^{208}$Pb via parity-violating electron scattering (PREX experiment) \cite{horo2001,roca2011} or by means of antiprotonic atom data \cite{brown2007,centelles2009} also constrain the density dependence of the symmetry energy because of the so-called Typel-Brown correlation  \cite{brown2006}. A recent update and a critical analysis of these constraints on the nuclear symmetry energy can be found in Ref. \cite{tsang2012}. Further details on these and other methods are given in other contributions to this special volume.

Additional information on the symmetry energy can be gathered from astrophysical 
observations of compact objects, which open a new window into both the bulk and 
 microscopic properties of nuclear matter at extreme isospin
 asymmetries \cite{lattimer2005}. In fact, the symmetry energy
 determines to a large extent the composition of $\beta$-stable
 matter and therefore the structure and mass of neutron
 stars \cite{schulze2006}. In particular, the characterization of the
 core-crust transition in neutron stars \cite{pieka2001,xu2009,mou10,ducoin2010},
 or the analysis of power-law correlations, such as the relation between the radius 
of a neutron star and the equation of state \cite{lat01},
can put stringent constraints on the symmetry energy.
From the theoretical point of view, the symmetry energy 
has been determined using both phenomenological and microscopic many-body approaches.
Phenomenological approaches, either relativistic or non-relativistic, are based on 
effective interactions that are usually fit to reproduce the binding
energy of stable nuclei \cite{stone2007}. 
Therefore,  predictions at high asymmetries should be taken with care.
The Skyrme-Hartree-Fock \cite{flocard1978} and the Relativistic Mean 
Field \cite{serot1986} methods are the most common ones. 
However, in spite of the large amount of constraints imposed in the fitting procedures of 
the effective interactions, there is still a large dispersion on the results
for the symmetry energy (and its derivatives) provided by the phenomenological approaches. 
Hence, fully microscopic approaches look as a safe and necessary alternative. 

Microscopic approaches  start from realistic nucleon-nucleon (NN) interactions 
that reproduce the scattering and bound state properties of the free two-nucleon system.
In-medium correlations are then built using many-body 
techniques that  incorporate the effects of the nuclear medium and account for isospin
asymmetry effects such as, for instance, the difference in the Pauli blocking
factors of neutrons and protons in asymmetric matter \cite{bombaci1991}.
Among this type of approaches the most popular ones are the
Brueckner--Bethe--Goldstone (BBG) \cite{day1967} and the Dirac--Brueckner--Hartree--Fock 
\cite{dbhf} theories, the variational method \cite{variational}, the correlated basis function
formalism \cite{cbf}, the self--consistent Green's function technique
(SCGF) \cite{scgf} or, recently, perturbative calculations using  V$_{low k}$ interactions \cite{vlowk}. 
In this work, we discuss results for the Brueckner--Hartree--Fock (BHF) \cite{day1967} approximation of the 
BBG theory and for the SCGF approach.

Unfortunately, whatever realistic two-nucleon force \linebreak (2NF) is used in a non-relativistic 
many-body calculation, the saturation properties of nuclear matter fail to be reproduced. 
Saturation densities are too large and saturation energies 
too attractive, with  calculations falling in the so-called Coester band \cite{coester}.  
Three-body forces (3BF) are expected to take care of this limitation. 
3BFs are also required in light nuclei,  
whose binding energies are not correctly predicted when computed with 2NF only \cite{wiringa2002}.
In this work, we employ the Argonne V18 (AV18) NN potential \cite{stoks1995} in all calculations.
Moreover, BHF calculations have been supplemented with the Urbana 
IX 3BF, reduced to a two-body density-dependent force by averaging over the third nucleon in the medium \cite{ferreira1999}. 
The extension of the SCGF formalism to include 3BFs has been achieved only recently \cite{arianna2013}.

In the following, we report microscopic calculations of the nuclear symmetry energy and its density dependence \cite{vida2009}. We also explore the different effect of NN correlations on SNM
 and PNM. We discuss how the isospin dependence of NN correlations affects the symmetry energy. 
To this end, we study the contribution of the different terms in the NN interaction, particularly the tensor one, to the symmetry energy \cite{vida2011}.
We describe how NN correlations produce high-momentum components and how these affect the kinetic energy and the symmetry energy \cite{rios2012}.  
As mentioned above, 
the calculations are performed in the framework of BHF and SCGF approaches, which are well suited for this 
type of analysis.    

\begin{figure}
\resizebox{0.45\textwidth}{!}
{%
\includegraphics[clip=true]{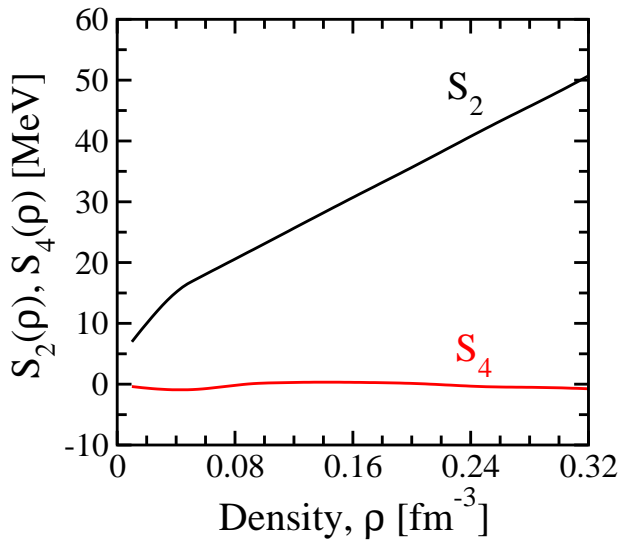}
}
\caption{(Color online) Density dependence of the symmetry energy coefficients $S_2$ and $S_4$
calculated in the BHF approximation using the AV18  interaction plus a 3BF of the Urbana type, as indicated  in the text.}
\label{fig:1}       
\end{figure}

\begin{figure*}
\begin{center}
\resizebox{0.75\textwidth}{!}
{%
\includegraphics[clip=true]{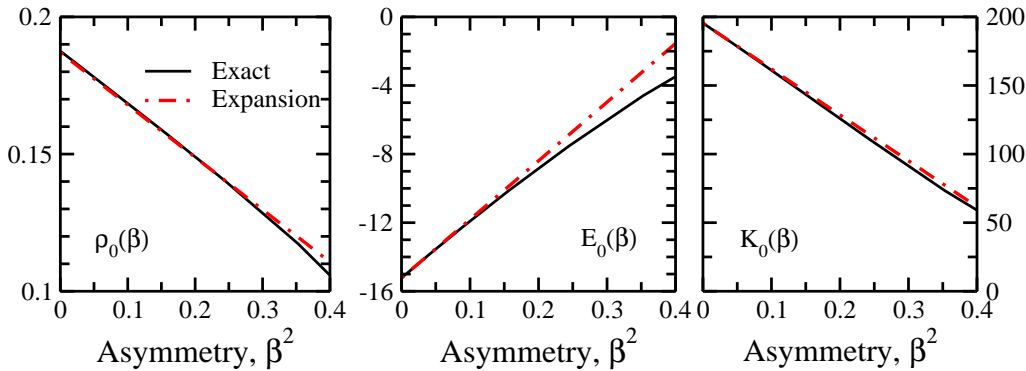}
}
\caption{(Color online) Isospin asymmetry dependence of the density (left panel), energy per particle (middle panel), and incompressibility coefficient (right panel) at the saturation point of asymmetric nuclear matter. Solid lines show the results of the BHF calculation whereas dashed lines indicate the results of the expansion of Eqs.\ (\ref{eq12a}), (\ref{eq12b}) and (\ref{eq12}). Units of 
$E_0(\beta)$ and $K_0(\beta)$ are given in MeV, whereas $\rho_0(\beta)$ is given in fm$^{-3}$.}
\label{fig:2}       
\end{center} 
\end{figure*}


 \section{Isospin asymmetric nuclear matter}
\label{sec:1}
Assuming charge symmetry for nuclear forces, the energy per particle of asymmetric
 nuclear matter can be expanded around SNM in the isospin asymmetry parameter, 
$\beta=(N-Z)/(N+Z)=(\rho_n-\rho_p)/\rho$ only in terms of even powers of $\beta$:
\begin{equation}
\frac {E}{A} (\rho,\beta) = E_{SNM}(\rho) + S_2(\rho) \beta^2 + S_4(\rho) \beta^4 +
{\cal O}(6) \, .
\label{eq1}
\end{equation}
Here, $E_{SNM}(\rho)$ is the energy per particle of SNM, $S_2(\rho)$ is 
identified (neglecting  surface contributions \cite{dani2009,centelles2009}) with the usual symmetry energy 
in the semiempirical mass formula
\begin{equation}
S_2(\rho) = \frac {1}{2} \frac {\partial^2 E/A}{\partial \beta^2 } \Big |_{\beta=0} \, ,
\label{eq2}
\end{equation}
and 
\begin{equation}
S_4(\rho) = \frac {1}{24} \frac {\partial^4 E/A}{\partial \beta^4} \Big |_{\beta=0} \, .
\label{eq3}
\end{equation}
The dominant dependence of the energy per particle of asymmetric nuclear matter on $\beta$
 is essentially quadratic \cite{bombaci1991,lee98,frick2005}.
 Therefore, contributions from $S_4$ and other higher
 terms can be neglected. One can then estimate the symmetry energy by subtracting
 the energy per particle of PNM and that of SNM, according to 
\begin{equation}
S_2(\rho) \sim \frac {E}{A}(\rho,1) - \frac {E}{A}(\rho,0) \, .
\end{equation}
To check this approximation, we plot in Fig.\ \ref{fig:1} the density dependence of 
the coefficients $S_2$ and $S_4$ obtained  in our BHF calculation. As expected, the 
coefficient $S_4$ is very small  
and $S_2(\rho)$ is an increasing function of $\rho$ in the density region considered
($0-0.3$ fm$^{-3}$). In other words, the energy per particle of PNM is always larger
than that of SNM and no isospin instability shows up \cite{engvik1997}.  

To characterize the density dependence of the symmetry energy around 
saturation, it is useful to perform  a series expansion in terms of the density. 
To this end, one considers first the density dependence of the energy per 
particle of SNM around the saturation density
 $\rho_0$ in terms of a few bulk parameters, 
\begin{equation}
E_{SNM} (\rho) = E_0 + \frac{K_0}{2} \left ( \frac {\rho-\rho_0}{3 \rho_0} \right )^2 
+ \frac {Q_0}{6} \left ( \frac {\rho-\rho_0}{3 \rho_0}\right )^3 + {\cal {O}}(4) \, .
\label{eq5}
\end{equation}
The coefficients $E_0$, $K_0$ and $Q_0$ correspond to the energy per particle, the incompressibility coefficient, and  the
 third derivative of the energy of SNM at saturation, 
\begin{equation}
E_0 = E_{SNM}(\rho=\rho_0), ~~~~ K_0 = 9 \rho_0^2 \frac {\partial^2 E_{SNM}(\rho)}{\partial \rho^2 } \Big |_{\rho=\rho_0} \, , 
\end{equation}
and 
\begin{equation}
Q_0 = 27 \rho_0^3 \frac {\partial^3 E_{SNM}(\rho)}{\partial \rho^3}  \Big |_{\rho=\rho_0} \, .
\label{eq7}
\end{equation}

Similarly, the symmetry energy around saturation can also be characterized in
 terms of  a few parameters, 
\begin{eqnarray}
S_2(\rho) &=&  E_{sym} + L \left ( \frac {\rho-\rho_0}{3 \rho_0} \right ) +
\frac {K_{sym}}{2} \left ( \frac {\rho-\rho_0}{3 \rho_0 }\right)^2   \nonumber \\
          &+& \frac {Q_{sym}}{6} \left ( \frac {\rho- \rho_0}{3 \rho_0}\right ) ^3 + {\cal O}(4) \, ,
\label{eq8}
\end{eqnarray}
where $E_{sym}$ is   the symmetry energy at saturation, and the quantities 
$L$, $K_{sym}$ and $Q_{sym}$ are related to its slope, curvature, and third derivative, 
at saturation density,
\begin{equation}
L= 3 \rho_0 \frac {\partial S_2(\rho)}{\partial \rho}\Big | _{\rho=\rho_0}, ~~~
K_{sym} = 9 \rho_0^2 \frac {\partial^2 S_2(\rho)}{\partial \rho^2 } \Big |_{\rho=\rho_0}
\, ,
\label{eq:L}
\end{equation}
and
\begin{equation}
Q_{sym} = 27 \rho_0^3 \frac { \partial^3 S_2(\rho)}{\partial \rho^3 }\Big |_{\rho=\rho_0} \, .
\label{eq10}
\end{equation}

Combining the expansions of Eqs.~(\ref{eq1}), (\ref{eq5}) and (\ref{eq8}), one can predict the existence of a saturation
 density, satisfying a zero pressure condition. For a given asymmetry, the energy per particle can be expanded around the new, asymmetry-dependent saturation density, $\rho_0(\beta) \sim \rho_0 ( 1 - 3 (L/K_0) \beta^2 )$, as 
\begin{eqnarray}
\frac {E}{A}(\rho,\beta) &=& E_0(\beta) + \frac {K_0(\beta)}{2} \left (
\frac {\rho-\rho_0(\beta)}{3 \rho_0(\beta)}\right )^2  \nonumber \\
           & + & \frac {Q_0(\beta)}{6} \left ( \frac {\rho -\rho_0(\beta)}{3 \rho_0(\beta)} \right )^3 + {\cal O}(4) \, ,
\label{eq11}
\end{eqnarray}
where the coefficients $E_0(\beta)$, $K_0(\beta)$, and $Q_0(\beta)$ define
 the energy per particle, the incompressibility coefficient, and the third
 derivative at 
the new saturation density, $\rho_0(\beta)$. These coefficients can be written in terms
of the quantities defined at $\rho_0$, {\it i.e.,} the saturation density for $\beta=0$: 
\begin{equation}
E_0(\beta) = E_0 + E_{sym} \beta^2 + {\cal O}(4)  \, ,
\label{eq12a}
\end{equation}
\begin{equation}
K_0(\beta) = K_0 + ( K_{sym} - 6 L - \frac{Q_0}{K_0} L ) \beta^2 + {\cal O}(4)  \, ,
\label{eq12b}
\end{equation}
\begin{equation}
Q_0(\beta) = Q_0 + ( Q_{sym} - 9 L \frac {Q_0}{K_0} ) \beta^2  + {\cal O}(4) 
\, .
\label{eq12}
\end{equation}

In Fig.~\ref{fig:2},  we explore the behavior  of the saturation density $\rho_0(\beta)$
(left panel), energy per particle $E_0(\beta)$ (middle panel) and incompressibility
$K_0(\beta)$ (right panel) as a function of  $\beta^2$, up to an asymmetry 
of $\beta \sim 0.6$, for which one still finds a saturation density. The figure,
shows  the very good agreement between the expansion up to second order
in $\beta$ (dashed lines) and the exact numerical calculations (solid lines). 
For $\beta=0$, one recovers the results of SNM. As $\beta$ increases, however,
the saturation density, the binding energy, and the incompressibility decrease. 

\begin{figure}
\begin{center}
\resizebox{0.400\textwidth}{!}
{%
\includegraphics[clip=true]{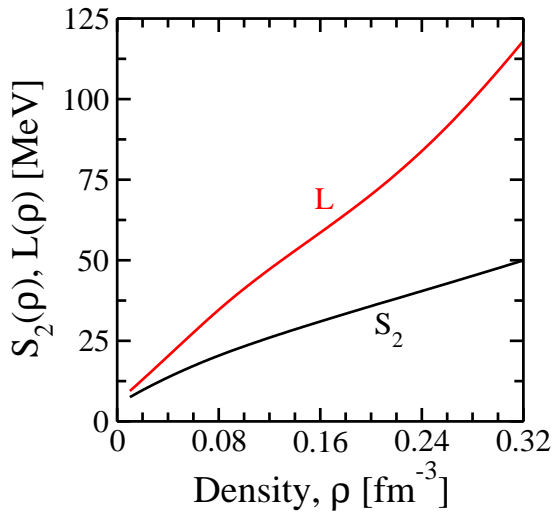}
}
\caption{(Color online) Density dependence of the symmetry energy and the slope parameter $L$
calculated in the BHF approximation using the AV18  interaction plus a 3BF of Urbana type as 
indicated  in the text.}
\label{fig:3} 
\end{center}
\end{figure}

We finish this section by showing in  Fig.\ \ref{fig:3} the density dependence of the symmetry energy $S_2(\rho)$ 
and its slope parameter, $L$, defined in Eq.~(\ref{eq:L}), as obtained in our BHF calculation. 

\section{The tensor component of the NN interaction}
\label{sec:2}
Realistic NN interactions should fulfill a minimum set of requirements.
 In particular, realistic potentials are built to reproduce the Nijmegen database
 \cite{stoks1993}, which contains a full set of NN elastic scattering phase shifts up to energies
 of about 350 MeV, with and accuracy of $\chi^2/N_{data} \sim 1$. 
Only potentials that fulfill this condition should be used  as input for
 the so-called {\it ab initio} many-body schemes, which aim at providing
 a {\it first-principles} description of the equation of state (EoS)  of PNM 
and SNM. 
The Argonne V18 potential \cite{stoks1995} is one such realistic interaction, that has been used
for {\it ab initio} calculations in  nuclear matter and finite nuclei with a diversity
of many-body approaches \cite{baldo2012}. 

Applying symmetry arguments, the  strong interaction part of the AV18   potential can be  expressed
 as a sum of 18 operators,
\begin{equation}
V_{ij} = \sum_{p=1,18} v_p(r_{ij}) O_{ij}^p \, .
\end{equation} 
The first 14 operators are associated to the spin, isospin, tensor, spin-orbit, and quadratic spin-orbit components of the nuclear force: 
\begin{eqnarray}
O_{ij}^{p=1,...,14}  &=& 1, { \vec \tau}_i \cdot { \vec \tau}_j, { \vec \sigma}_i \cdot { \vec \sigma}_j, 
({ \vec \tau}_i \cdot  { \vec \tau}_j)(  { \vec \sigma}_i \cdot  {\vec  \sigma}_j),\nonumber \\
  & & S_{ij}, S_{ij} ({\vec  \tau}_i \cdot  {\vec  \tau}_j), {\bf L}\cdot {\bf S}, {\bf L}\cdot {\bf S} 
({\vec  \tau}_i \cdot  {\vec  \tau}_j), \nonumber \\
  & & L^2, L^2 ({\vec  \tau}_i \cdot {\vec  \tau}_j), L^2 ({\vec  \sigma}_i \cdot {\vec  \sigma}_j), \nonumber \\
& & L^2 ({\vec  \tau}_i \cdot  {\vec  \tau}_j)
({\vec  \sigma}_i \cdot {\vec  \sigma}_j), \nonumber \\
 & & ({\bf L}\cdot {\bf S})^2, ({\bf L}\cdot {\bf S})^2 ({\vec  \tau}_i \cdot {\vec  \tau}_j) \, .
\label{eq:pot}
\end{eqnarray}
The four additional operators,
\begin{equation}
O_{ij}^{p=15, ..,18} = T_{ij},~ T_{ij}({\vec  \sigma}_i {\vec  \sigma}_j),~ T_{ij} S_{ij},~ (\tau_{zi}+\tau_{zj}) \, ,
\end{equation}
where $T_{ij}=  3 \tau_{zi}\tau_{zj} - {\bf \tau}_i {\bf \tau}_j$ is the isotensor operator, break charge
 independence. The radial functions that multiply each operator are adjusted by fitting experimental
 data on two-body scattering phase shifts as well as deuteron properties.

\begin{table}
\caption {Deuteron  D-state probability $P_D$, quadrupole moment $Q_d$ (in fm$^2$),
 total binding energy, kinetic and potential energy, and their decomposition (in the second row)
 in partial waves,
 calculated for the AV18 NN interaction. All energies are given in MeV.}
\label{tab:1}
\begin{center}
\begin{tabular}{lllll}
\noalign{\smallskip}\hline
    $P_D (\%)$    & $Q_d$  & $E$ & $K$  & $V$ \\
\hline\noalign{\smallskip}
\hline\noalign{\smallskip}
 $5.78$  & $0.27$  &$ -2.24$  & $19.86$  & $-22.10$ \\
\hline\noalign{\smallskip}
   &   &  &   &  \\
\hline\noalign{\smallskip}
    $K_S$  &  $K_D$  & $V_S$  &  $V_D$  &  2 $V_{SD}$  \\
\hline\noalign{\smallskip}
\hline\noalign{\smallskip}
 $11.30$ & $8.56$ & $-3.95$ & $0.77$ & $-18.91$ \\
\hline\noalign{\smallskip}
\end{tabular}
\end{center}
\label{tab1}
\end{table}

The tensor component of the NN interaction, which is crucial in the generation 
of NN correlations in the nuclear medium,  plays a central  role in the
 reproduction of  the experimental phase shifts. Moreover, the tensor 
is also largely responsible for the structure  and binding  energy of the deuteron, the simplest  bound   nucleonic system. 
Even though the deuteron only 
probes the NN potential in the $^3S_1-^3D_1$ partial waves, the analysis of the
independent contributions of the different waves ($S-$, $D-$ and mixed channel) 
highlights the importance of the different components of the NN interaction \cite{baldo2012,polls1998}.
The first observation is that for a realistic potential, as AV18, the binding energy of
the deuteron results from a strong cancellation 
between the kinetic  and the interaction energies (see first row of Table \ref{tab1}). For the AV18, the  binding energy,
$E=-2.24$ MeV, comes from a large kinetic, $K=19.86$ MeV,  and  interaction, $V= -22.10$ MeV, energies.
Note that this is the binding energy  obtained  with only the strong interaction part of AV18, {\it i.e.,} 
when all small 
electromagnetic terms are omitted. These
repulsive electromagnetic terms shift the binding energy to the true experimental value $E=-2.22$ MeV. 
It is also worth stressing that the charge-dependent terms of
 $V_{p} (p=15, ...,18)$, described in terms of an isotensor operator, have no contribution in the isosinglet
 deuteron state.

The $D-$state probability as well as the quadrupole moment are also a 
direct consequence of the tensor 
component of the nuclear force that allows for the coupling between
 the $S$ and the $D$ wave. 
It is instructive  to separate the contributions of the $^3S_1$ and $^3D_1$ states
 to the total kinetic and potential  energies (see second row of Table \ref{tab1}). Assuming that the deuteron is a properly
 normalized combination of the
$^3S_1$ and $^3D_1$ partial waves,  we define the contributions of the $S$ and $D$ states to 
the kinetic energy, $K_S = \langle ^3S_1 | K | ^3S_1\rangle $ and 
$K_D= \langle ^3D_1 | K | ^3D_1 \rangle$, and to the potential
 energy, $V_S =  \langle ^3 S_1 | V| ^3S_1\rangle$ and 
$V_D = \langle ^3D_1 | V | ^3D_1\rangle $. 
The latter also receives a contribution from the $^3S_1-^3D_1$
 mixing, $V_{SD}= \langle ^3S_1 | V | ^3D_1 \rangle$.
 Actually the largest contribution 
comes from the mixing term, $V_{SD}$, which  accounts  for more than $85 \%$ of the
 final value of the potential  energy. 
 
\begin{table} 
\caption { Contribution of the different components of the NN AV18 interaction to the binding
 energy of the deuteron. All energies are given in MeV.}
\begin{center}
\begin{tabular}{llll}
\hline\noalign{\smallskip}
    Central     & Tensor   & Spin-orbit  & $L^2$  \\
\hline\noalign{\smallskip}
\hline\noalign{\smallskip}
 -4.45   & -16.62   & -3.75   & 2.72  \\
\noalign{\smallskip}\hline
\label{tab:2}
\end{tabular}
\end{center}
\end{table}

Additional information can be obtained by
 looking at  the expectation values of the different components of the potential  on
 the ground-state wave function of the deuteron. We have grouped the 18 components
 into four sets: the first four operators ($p=1, \ldots ,4$), the tensor components, 
 $S_{ij}$ ($p=5,6$), the spin-orbit components ($p=7, 8$ and $p=13, 14$), and the quadratic
 orbital angular momentum 
components $L^2$ ($p=9, \ldots ,12$). The group of charge-dependent terms, $p=15, \ldots ,18$, 
does not contribute to the deuteron. The results of this decomposition are presented 
in Table \ref{tab:2}. As expected, the largest contribution corresponds to the tensor
 component. All contributions are attractive, except the one proportional to $L^2$, which
 is slightly repulsive. Notice also that the spin-orbit contribution is 
non-negligible and amounts to $17 \, \%$ of the total potential energy.

\section{Brueckner--Hartree--Fock results}
\label{sec:3}
The energy per particle of asymmetric nuclear matter in the BHF approach is calculated as 
\begin{equation}
\frac {E}{A}(\rho, \beta) = \frac {1}{A} \sum_{\tau} \sum_{\mid k\mid < k_{F_{\tau}}} 
\left ( \frac {\hbar^2 k^2}{2 m} + \frac {1}{2} Re \left [U_{\tau(k)} \right ] \right ) \, ,
\end{equation}
where $U_{\tau}(k)$ represents the mean field ``felt'' by a nucleon ($\tau =n,p$) due to 
its interaction with the other nucleons of the medium. 
$U_{\tau}(k)$ is calculated through the ``on-shell energy'' $G$ matrix,
\begin{equation}
U_{\tau} = \sum_{\tau'} \sum_{\mid k\mid < k_{F_{\tau'}}} \langle \vec k \vec k' \mid 
G_{\tau \tau'; \tau \tau'}(\omega=\epsilon_{\tau}(k)+\epsilon_{\tau'}(k')) \mid \vec k \vec k' \rangle_A \, ,
\end{equation}
where the sum runs over all neutron and proton occupied states, and the matrix elements are properly antisymmetrized.
The single particle energy, $\epsilon_{\tau}(k)$,  is defined in terms of the single-particle kinetic energy and 
the single-particle potential,  
\begin{equation}
\epsilon_{\tau}(k) = \frac {\hbar^2 k^2}{2 m_{\tau}} + Re[U_{\tau}(k)] \, .  
\end{equation}
Finally, the $G$-matrices describing the effective interaction between two nucleons in 
the medium are constructed by solving  the Bethe-Goldstone equation
\begin{eqnarray}
G_{\tau_1 \tau_2; \tau_3 \tau_4} (\omega ) &=& V_{\tau_1 \tau_2; \tau_3 \tau_4} \\ \nonumber + \sum_{jk} 
                       &  V&_{\tau_1 \tau_2; \tau_j \tau_k} 
                                        \frac {Q_{\tau_j \tau_k}}{\omega -\epsilon_j -\epsilon_k + i \eta } G_{\tau_j \tau_k; \tau_3 \tau_4} (\omega)  \, , 
\end{eqnarray}
where $V$ denotes the free-space NN interaction, $Q_{\tau_j \tau_k}$ is the Pauli operator  which  allows only for intermediate states compatible  with the 
Pauli principle, and $\omega$ is the so-called starting energy, which corresponds to the sum of non-relativistic
energies of the interacting nucleons.  Note that the whole procedure requires a self-consistent process.  It has been shown by Song {\it et al.,} \cite{baldo1998} 
that the contribution to the energy from three-hole-line diagrams (that account for the effect of three-body correlations) is minimized when the so-called continuous 
prescription \cite{jeu76} is adopted for the in-medium potential, which is a strong
indication of the convergence of the BBG expansion.
We adopt this prescription in our calculation.

The BHF calculations  discussed in this work have been performed with the realistic AV18 
NN interaction supplemented with a 3BF of Urbana type. 
This 3BF contains two parameters that are fixed by requiring that the BHF calculation reproduces
 the energy and saturation density of SNM. 
The results reported here correspond to the original set of parameters 
of Baldo and Ferreira in Ref.~\cite{ferreira1999}. 
Moreover, the 3BF has been reduced to a two-body density-dependent force by averaging over the third nucleon 
in the medium \cite{ferreira1999}. See also Refs.~\cite{schulze2004,schulze12008,schulze22008}
for an extensive analysis of the use of 3BFs in nuclear and neutron matter.

\begin{table}
\caption {Bulk parameters characterizing the density dependence of the energy 
of SNM and the symmetry energy around the saturation density for our 
BHF calculation with and without 3BF. All quantities are in MeV, except $\rho_0$, given in 
fm$^{-3}$. }
\label{tab:3a}
\begin{center}
\begin{tabular}{lllll}
\noalign{\smallskip}\hline
        & $\rho_0$   & $E_0$  & $K_0$  & $Q_0$ \\
\hline\noalign{\smallskip}
\hline\noalign{\smallskip}
 BHF (no 3BF)   & 0.240  &  -17.30   & 213.6   & -225.1 \\
 BHF (3BF)      & 0.187  &  -15.23   & 195.5   & -280.9 \\

\hline\noalign{\smallskip}
                &  $E_{sym}$  & $L$  &  $K_{sym}$   &   $Q_{sym}$  \\
\hline\noalign{\smallskip}
\hline\noalign{\smallskip}
 BHF (no 3BF)   & 35.8  & 63.1  & -27.8  & -159.8    \\
 BHF (3BF)      & 34.3  & 66.5  & -31.3  & -112.8    \\
\hline\noalign{\smallskip}
\end{tabular}
\end{center}
\end{table}

We start the  discussion of the BHF results by  showing in Table \ref{tab:3a} the bulk parameters characterizing the
 density dependence  of the energy of SNM and the symmetry energy around saturation density.
 We report the BHF results obtained with and without three-body forces. 
 The comparison of the different quantities is strongly influenced by the fact that they 
are calculated at a value of the saturation density which is different with and without 3BF. 
Note that, in general, the effects of the 3BF are more important on the isoscalar properties, like $K_0$. 
Our BHF calculation gives a value of $L= 66.5$ MeV, compatible with recent experimental constraints (see {\it e.g.,} Fig. 1 of Ref.~\cite{vinas12}).

The properties associated with the density dependence of the symmetry energy are little affected by 
the 3BF. This is due to both a shift in saturation density and to a similarly repulsive effect on 
the energy of both SNM and PNM once the density shift is taken into account. 
Overall, there is a small dependence of the isovector properties on the 3BF, even though the independent 
contributions on SNM and PNM are not small. Recently, the importance of 3BFs has been revisited in 
the context of chiral perturbation theory \cite{ekstrom13}. New fitting protocols of two-nucleon 
forces seem to indicate that the effect of 3BF could actually be rather small in PNM. Nevertheless, the 
relative importance of the two- and three-body contributions change with the resolution scale. It is not 
clear that such observations, valid for the somewhat soft chiral interactions, apply when considering a hard 
interaction like AV18. In any case, one does not expect isovector properties to depend much on the presence or absence of 3BFs.

In the BHF approach,  one calculates the correction, $\Delta E_{BHF}$, to the energy of the 
free Fermi gas (FFG), {\it i.e.,} the underlying non-interacting system, and expresses the total energy as 
\begin{equation}
E_{BHF}= E_{FFG}~+~\Delta E_{BHF} \, .
\end{equation}
In  Table \ref{tab:3}, we  report  this decomposition for the energy 
per particle of SNM and PNM,  
at the saturation density provided by the AV18+3BF calculation.  The symmetry energy is calculated as 
  the difference of the total energy per particle of PNM and SNM. 
The FFG energy is larger for neutron matter than for symmetric matter 
and therefore its contribution  to the symmetry energy is  positive and 
amounts to $\sim 14.38$ MeV. $\Delta E_{BHF}$ is less attractive for neutron matter than for 
nuclear matter and also gives a positive contribution ($\sim 19.92$ MeV) to the  symmetry energy.
The addition of these two quantities, which are  of the same order, provides a symmetry energy 
of $\sim 34.3$ MeV. The contributions to $L$ can be decomposed similarly, but in this case 
the contribution of  $\Delta E_{BHF}$ (37.22 MeV) is slightly larger than 
that of the FFG, which amounts to 28.78 MeV.

\begin{table} 
\caption { Free-Fermi gas contribution, $\Delta E_{BHF}$ and  total energy per particle of 
PNM and  SNM. 
The respective contributions to $E_{sym}$ and $L$ are also reported. 
The results correspond to the 
saturation density of the BHF approach, $\rho_0 = 0.187$ fm$^{-3}$ for the AV18+3BF. All results
 are given in MeV}
\begin{center}
\begin{tabular}{lllll}

\hline\noalign{\smallskip}
    & $E_{PNM}$       & $E_{SNM}$    & $E_{sym}$   & $L$   \\
\hline\noalign{\smallskip}
\hline\noalign{\smallskip}
 FFG & 38.911    &  24.529    &  14.382    & 28.779  \\
 $\Delta E_{BHF}$ & -19.682 & -39.600 & 19.918 & 37.721 \\
 Total  & 19.229 & -15.071 & 34.300 & 66.500 \\
\noalign{\smallskip}\hline
\label{tab:3}
\end{tabular}
\end{center}
\end{table}

To get a further physical insight into $\Delta E_{BHF}$, it is useful to look at 
its  spin-isospin $(S,T)$ decomposition, reported 
in Table \ref{tab:4}.  As expected, the main contribution is from  the $(1,0)$ channel which is acting 
only in SNM and has a large attractive contribution. It is precisely in this
channel where the tensor component of the NN force is active.  Note  that the $T=1$ channels give
 similar contributions in nuclear and neutron matter and therefore its contribution to 
the symmetry energy is small. The channel (0,0)  gives 
a repulsive contribution to the total energy in SNM and since it does not play any 
role  in neutron matter, its contribution to the total symmetry energy is negative. Notice again that 
the tensor force is not acting in this channel.  

\begin{table}
\caption{ Spin-isospin (S,T) channel decomposition of $\Delta E_{BHF}$ for PNM 
and SNM. The respective contributions to $E_{sym}$ and $L$ are also reported.
The results correspond to the
saturation density of the BHF approach, $\rho_0 = 0.187$ fm$^{-3}$ for the AV18+3BF. All results 
 are given in MeV}
\begin{center} 
\begin{tabular}{lllll}
\hline\noalign{\smallskip}
 (S,T)  & $E_{PNM}$   & $E_{SNM}$ & $E_{sym}$ & $L$   \\
\hline\noalign{\smallskip}
\hline\noalign{\smallskip}
 (0,0)   &  0  & 5.894  & -5.894   & -23.085   \\
 (0,1) & -21.041  & -17.764  & -3.277   &-3.142   \\
 (1,0)  &  0 & -28.363   & 28.363   & 51.696   \\
 (1,1)  &  1.359 & 0.633  & 0.726  & 12.252  \\ 
\noalign{\smallskip}\hline
\label{tab:4}
\end{tabular}
\end{center}
\end{table}

Let us further this analysis by looking at the contributions of the different partial 
waves to $\Delta E_{BHF}$, as shown in Table \ref{tab:5}.
Notice that the $^1S_0$ contribution, which is dominated by the central component of the NN potential, 
has a similarly large contribution to both PNM and SNM and therefore 
its effect on the symmetry energy is almost negligible. 
The largest contribution  is provided by the $^3S_1-^3D_1$ partial wave, which corresponds
to $T=0$, active only in nuclear matter. For larger values of $J$, the contributions become smaller and 
many cancellations take place. 
In general, one observes that the final energy is the result of a large  cancellation between
$E_{FFG}$ and $\Delta E_{BHF}$  and that the absolute value of the correction $\Delta E_{BHF}$ 
for neutron matter is  significantly smaller than for nuclear matter. This observation points to the
 well accepted fact that neutron matter is less correlated than nuclear matter.

\begin{table}
\caption{Partial wave decomposition of $\Delta E_{BHF}$ for  PNM and SNM. 
The contributions of each partial wave to $E_{sym}$ and $L$ are also reported. 
The results correspond to the
saturation density of the BHF approach, $\rho_0 = 0.187$ fm$^{-3}$ for the AV18+3BF. All 
results are given in MeV.}
\label{tab:5}       
\begin{center}
\begin{tabular}{lllll}
\hline\noalign{\smallskip}
Partial wave  & $E_{PNM}$   & $E_{SNM}$ & $E_{sym}$ & $L$   \\
\hline\noalign{\smallskip}
\hline\noalign{\smallskip}
$^1S_0$ & -14.330 & -14.407 & 0.077 & 11.229 \\
$^3S_1$ & 0 & -24.865  & 24.865  & 35.521 \\
$^1P_1$ & 0 & 5.193  & -5.193 & -20.201  \\
$^3P_0$ & -4.522  & -3.713 & -0.809  & 0.224  \\
$^3P_1$ & 18.459  & 12.002  & 6.457  & 27.702  \\
$^3P_2$ & -13.550  & -8.102  & -5.448 & -17.784  \\
$^1D_2$ & -5.850  & -3.154 & -2.696 & -10.888 \\
$^3D_1$ & 0 & 1.036  & -1.036  & -3.894  \\
$^3D_2$ & 0 & -3.795  & 3.795  & 15.844 \\
$^3D_3$ & 0 & -0.522 & 0.522 & 3.305  \\
$^1F_3$ & 0 & 0.699 & -0.699 & -3.394 \\
$^3F_2$ & -0.651  & -0.221 & -0.430& -1.515 \\
$^3F_3$ & 2.022 & 0.826 & 1.196 & 5.026  \\
$^3F_4$ & -0.743 & -0.183 & -0.560  & -3.006 \\
$^1G_4$ & -0.810 & -0.247 & -0.563 & -3.029 \\
$^3G_3$ & 0      & 0.002 & -0.002 & 0.425 \\
$^3G_4$ & 0 & -0.213 & 0.213 & 0.449 \\
$^3G_5$ & 0 & -0.053 & 0.053 & 0.617 \\
$^1H_5$ & 0 & 0.029& -0.029 & 0.122 \\
$^3H_4$ & 0.034 & 0.040 & -0.007 & 0.224 \\
$^3H_5$ & 0.226 & -0.033 & 0.258 & 0.949 \\
$^3H_6$ & 0.044 & 0.035 & 0.010 & 0.136 \\
\noalign{\smallskip}\hline
\end{tabular}
\end{center}
\end{table}
  
In the case of the deuteron, the total  binding energy is the result of a strong cancellation 
between the  kinetic energy and the potential  energy. The large kinetic energy is 
a consequence  of the  NN correlations existing in the deuteron, {\it i.e.,} in the $^3S_1-^3D_1$ channel.
In that sense, we would like to study the  decomposition of the total energy 
of the infinite system in the kinetic and potential energy. This kinetic energy will contain 
the effects of correlations and will be larger than the energy of the FFG. Therefore  
the difference between the {\it correlated kinetic energy} and  $E_{FFG}$ can quantify
NN correlations.     

Unfortunately,  the BHF approach does not give direct access to the separate contribution of the
kinetic and potential energies because it does not provide
the correlated many-body wave function, $|\Psi\rangle$. However, it has been shown \cite{hf1} that the 
Hellmann--Feynman theorem \cite{hf2} can be used to 
estimate the ground state expectation value of the interaction energy.
The kinetic energy  can then be calculated simply by subtracting 
the expectation value of the  potential energy from $E_{BHF}$. 
Writing the nuclear hamiltonian as $H= T+ V$, and defining a 
$\lambda$ dependent potential,
$H(\lambda) = T + \lambda V $, the expectation value of the potential energy is given by:
\begin{equation}
 \langle V \rangle \equiv \frac {\langle \Psi | V | \Psi \rangle }{\langle \Psi | \Psi \rangle } = \left ( \frac {dE}{d\lambda } \right )_{\lambda=1} \, .
\label{eq15}
\end{equation}
In Table \ref{tab:6} we show the kinetic and potential energy contributions to the total 
energy of PNM, SNM,  $E_{sym}$ and $L$ at the saturation density, $\rho_0= 0.187$ 
fm$^{-3}$,  provided by the AV18+3BF within the BHF approach. 

\begin{table}
\caption { Kinetic, $\langle K \rangle$, and potential, $\langle V \rangle $, contributions to
 $E_{PNM}$, $E_{SNM}$,  $E_{sym}$ and $L$. Units are given in MeV.}
\label{tab:6}
\begin{center}

\begin{tabular}{lllll}

\hline\noalign{\smallskip}
       & $E_{PNM}$ & $E_{SNM}$ & $E_{sym}$ & $L$ \\
\hline\noalign{\smallskip}
\hline\noalign{\smallskip}
$ \langle K \rangle$ & 53.321 & 54.294 & -0.973 & 14.896 \\
$ \langle V \rangle$ & -34.251 & -69.524 & 35.273 & 51.604 \\
 Total & 19.070 & -15.230 & 34.300 & 66.500 \\
\noalign{\smallskip}\hline
\end{tabular}
\end{center}
\end{table}

As in the case of the deuteron, the total energy of both PNM and SNM are the
 result of a strong cancellation between the kinetic and potential energies.
It is worth noticing that the kinetic energy contribution to $E_{sym}$ is very small and
slightly negative. This is in contrast to the results for the FFG (see Table \ref{tab:3}). 
The increase of the kinetic energy with respect to the FFG energy, which is due mainly
 to short  range  and tensor correlations, is much larger for SNM  than for PNM. 
 Again this is an indication that, at the same density, SNM is more
 correlated than PNM. 

It is also worth mentioning that the kinetic contribution to $L$ is smaller than the  corresponding
 one of the FFG ($L_{FFG} \sim 29.2 $MeV) reported in Table \ref{tab:3}. Clearly the major 
contribution to both $E_{sym}$ and
 $L$ is due to the  potential energy part. Note that this contribution is
 practically equal to the  total value of $E_{sym}$ and it represents $\sim 78\% $ of $L$. Results along these lines 
have been recently reported by Xu and Li \cite{xu2011} using a phenomenological model for $n(k)$.

\begin{table}
\caption{Spin-isospin  (S,T) channel decomposition of the potential contribution 
to $E_{PNM}$, $E_{SNM}$ and $E_{sym}$ and $L$. The results correspond to the
saturation density of the BHF approach, $\rho_0 = 0.187$ fm$^{-3}$ for the AV18+3BF. All results 
 are given in MeV.}
\begin{center}
\begin{tabular}{lllll}
\hline\noalign{\smallskip}
 (S,T)  & $E_{PNM}$   & $E_{SNM}$ & $E_{sym}$ & $L$   \\
\hline\noalign{\smallskip}
\hline\noalign{\smallskip}
 (0,0)   &  0  & 5.6  & -5.6  & -21.457  \\
 (0,1) & -29.889  & -23.064 & -6.825  &-17.950  \\
 (1,0)  &  0 & -49.836  & 49.836  & 90.561  \\
 (1,1)  &  -4.362 & -2.224 & -2.138 & 0.450 \\ 
\noalign{\smallskip}\hline
\label{tab:7}
\end{tabular}
\end{center}
\end{table}

The spin-isospin ($S$, $T$) channel decomposition of the potential  part of $E_{PNM}$, $E_{SNM}$, $E_{sym}$ and $L$
is also illustrative. This  is reported  in Table \ref{tab:7} at $\rho_0$.
As in the case of $\Delta E_{BHF}$, the largest contribution to both $E_{sym}$ and $L$ is provided 
by the ($S=1$, $T=0$) channel, which is where the tensor is active. 
Interestingly, the $S=0$ channels have a small and similar negative contribution 
to $E_{sym}$ and also a moderate negative contribution to $L$ showing in total a strong cancellation 
with the contribution of the channel ($S=1$, $T=0$). However, the origin of these contributions is
qualitatively different. While the channel ($S=0$, $T=0$) does not contribute to neutron matter and 
has a small repulsive contribution to nuclear matter, the contribution of the  channel 
($S=0$,$T=1$) is the  result of a strong cancellation of large attractive contributions 
of this channel in both PNM and SNM. 
Analogous conclusions can be obtained from  Table \ref{tab:8}, where the partial wave decomposition 
of the potential  energy is reported.  Note that similar arguments have been already pointed out by other authors 
\cite{bombaci1991,engvik1997,xu2011,pandha1972,wiringa1988,rho1990,mue1987,zuo99,dieperink2003,li06,xu2010,li11,samarruca2011}.

Next, we analyze the contribution to the potential energy of the  different components of the
nuclear force. To such end, we apply the Hellmann--Feynman theorem to the separate
components of the AV18 potential and the Urbana IX three-body force.  The results 
are reported in Table \ref{tab:9}. The central contribution $\langle V_1 \rangle$ is large,
attractive and similar for neutron and nuclear matter and therefore gives a small contribution to
 $E_{sym}$. The largest contribution to isovector properties is from the tensor $\langle V_{S_{ij} (\vec \tau_i \vec \tau_j)}\rangle $,  which acts very efficiently to supply attraction in SNM.

\begin{table}
\caption{Partial wave decomposition of the potential part of $E_{PNM}$, $E_{SNM}$, $E_{sym}$ and $L$ . 
The results correspond to the
saturation density of the BHF approach, $\rho_0 = 0.187$ fm$^{-3}$ for the AV18+3BF.
Units are given in MeV.} 
\label{tab:8}
\begin{center}
\begin{tabular}{lllll}
\hline\noalign{\smallskip}
Partial wave  & $E_{PNM}$   & $E_{SNM}$ & $E_{sym}$ & $L$   \\
\hline\noalign{\smallskip}
\hline\noalign{\smallskip}
$^1S_0$ & -23.070 & -19.660 & -3.410 & -3.459 \\
$^3S_1$ & 0 & -45.810 & 45.810 & 71.855 \\
$^1P_1$ & 0 & 4.904 & -4.904 & -18.601 \\
$^3P_0$ & -5.312 & -4.029 & -1.292 & -1.898 \\
$^3P_1$ & 16.110 & 10.720 & 5.390 & 21.9149 \\
$^3P_2$ & -16.000 & -9.334 & -6.666& -21.168 \\
$^1D_2$ & -5.956 & -3.201& -2.755 & -11.033 \\
$^3D_1$ & 0 & 0.981 & -0.981 & -3.739 \\
$^3D_2$ & 0 & -3.982 & 3.982 & 16.601 \\
$^3D_3$ & 0 & -0.798 & 0.798 & 4.895 \\
$^1F_3$ & 0 & 0.694 & -0.694 & -3.348 \\
$^3F_2$ & -0.695 & -0.229 & -0.466& -1.799 \\
$^3F_3$ & 2.000 & 0.821 & 1.179 & 4.883 \\
$^3F_4$ & -0.796 & -0.194 & -0.602 & -3.239 \\
$^1G_4$ & -0.812 & -0.247 & -0.565 & -3.036 \\
$^3G_3$ & 0      & -0.001 & 0.001 & 0.441 \\
$^3G_4$ & 0 & -0.213 & 0.213 & 0.449 \\
$^3G_5$ & 0 & -0.057 & 0.057 & 0.650 \\
$^1H_5$ & 0 & 0.029& -0.029 & 0.107 \\
$^3H_4$ & 0.033 & 0.040 & -0.007 & 0.232 \\
$^3H_5$ & 0.225 & -0.033 & 0.258 & 0.968 \\
$^3H_6$ & 0.043 & 0.034 & 0.009 & 0.144 \\
\noalign{\smallskip}\hline
\end{tabular}
\end{center}
\end{table}

As mentioned above, the Urbana IX 3BF is reduced to an effective density-dependent two-body force when used in the BHF approach. This reduced, effective two-body force contains 
three components of the type $u_p(r_{ij},\rho) O_{ij}^p$ where 
$O_{ij}^{p=1,3} = 1, (\vec \sigma_i \cdot \vec \sigma_j) (\vec \tau_i \cdot \vec \tau_j),
 S_{ij} (\vec \tau_i \cdot \vec \tau_j)$. This introduces additional central, $\sigma \tau$ and tensor 
terms, which are reported on the last three rows of Table \ref{tab:9} (see {\it e.g.,} Ref.~\cite{ferreira1999} for details). 
The contribution of the two-body density dependent effective force to $E_{sym}$ can be
considered small, with the tensor component being the most important. 
These results clearly confirm that the tensor force gives the largest contribution to both $E_{sym}$ and $L$. 
The contributions from the other components are either negligible,  or almost cancel out. 
  
\begin{table}
\caption{Contributions to $E_{PNM}$, $E_{SNM}$, $E_{sym}$ and $L$ from the different components of the AV18 potential 
(indicated as $\langle V_i\rangle$) and the reduced Urbana three-body force (indicated as $\langle U_i\rangle$). The results correspond to the
saturation density of the BHF approach, $\rho_0 = 0.187$ fm$^{-3}$ for the AV18+3BF. Units are given in MeV.}
\label{tab:9}       
\begin{center}
\begin{tabular}{lllll}
\hline\noalign{\smallskip}
$\langle V \rangle$ & $E_{PNM}$   & $E_{SNM}$ & $E_{sym}$ & $L$   \\
\hline\noalign{\smallskip}
\hline\noalign{\smallskip}
 $\langle V_1\rangle$  & -31.212 & -32.710 & 1.498 & -5.580 \\
 $\langle V_{\vec \tau_i \vec \tau_j}\rangle$ & -4.957 & 3.997 & -8.954 &-20.383 \\
 $\langle V_{\vec \sigma_i \vec \sigma_j}\rangle $ & -0.319 & -0.382 & 0.063 & 2.392 \\
 $\langle V_{(\vec \tau_i \vec \tau_j) (\vec \sigma_i \vec \sigma_j)}\rangle $ & -5.724 & -11.388 & 5.664 & 2.521 \\
 $\langle V_{S_{ij}}\rangle $ & -0.792 & 1.912 & -2.704 & -4.998 \\
 $\langle V_{S_{ij} (\vec \tau_i \vec \tau_j)}\rangle $ & -4.989 & -37.592 & 32.603 & 47.095 \\
 $\langle V_{\vec L \vec S}\rangle $ & -7.538 & -1.754 & -5.784 & -12.251 \\
 $\langle V_{\vec L \vec S (\vec \tau_i \vec \tau_j)}\rangle $ &-2.671 & -6.539 & 3.868 & 3.969 \\
 $\langle V_{L^2}\rangle $ & 11.850 & 13.610 & -1.760 & 1.521 \\
 $\langle V_{L^2 (\vec \tau_i \vec \tau_j)}\rangle$ & -2.788 & 0.270 & -3.058 & -14.262 \\
 $\langle V_{L^2 (\vec \sigma_i \vec \sigma_j) }\rangle$ & 1.265 & 1.383 & -0.118 & 1.405 \\
 $\langle V_{L^2 (\vec \sigma_i \vec \sigma_j) (\vec \tau_i \vec \tau_j)}\rangle$ & 0.051 & 0.008 & 0.043 & -0.341 \\
 $\langle V_{(\vec L \vec S)^2}\rangle $ & 4.194 & 5.682 & -1.488 & -0.327 \\
 $\langle V_{(\vec L \vec S)^2 (\vec \tau_i \vec \tau_j)}\rangle$ & 5.169 & -6.190 & 11.359 & 31.368 \\
 $\langle V_{T_{ij}}\rangle $ & 0.003 & 0.039 & -0.036 & -0.022 \\
 $\langle V_{T_{ij} (\vec \sigma_i \vec \sigma_j)}\rangle$ & -0.017 & -0.106 & 0.089 & 0.042 \\
 $\langle V_{T_{ij} S_{ij}}\rangle$ & 0.004 & 0.079 & -0.075 & -0.124 \\
 $\langle V_{(\tau_{z_i}+\tau_{z_j})}\rangle $ & -0.084 & -0.001 & -0.083 & -0.331 \\
                               &        &        &        &         \\
 $ \langle U_1\rangle $     & 2.985 & 3.251 & -0.266 & -0.630 \\
 $\langle U_{ (\vec \tau_i \vec \tau_j) (\vec \sigma_i \vec \sigma_j)}\rangle$ & 2.252 & 3.999& -1.745 & -7.228\\ 
 $ \langle U_{S_{ij} (\vec \tau_i \vec \tau_j)}\rangle $ & -0.935 & -7.092 & 6.157 & 27.768 \\
\noalign{\smallskip}\hline
\end{tabular}
\end{center}
\end{table}

\section{High-momentum components in the symmetry energy}
\label{sec:4}
In the previous section, we have used 
the  kinetic energy difference between  the correlated system and  the underlying FFG
as a measure of NN correlations. The kinetic energy is the result of integrating the momentum distribution weighted with the 
kinetic energy associated to each momentum. The presence of correlations modifies the step
function associated to the FFG, $\Theta (k_F-k)$. There is a promotion of particles to higher
momentum states and, as a consequence, there is also a depletion below $k_F$. Thus, the single-particle 
momentum distributions can be taken also as a probe of the correlations embedded in the nuclear wave function. 

In this section, we would like to discuss the effects of correlations directly on the momentum distributions. 
To this end, we rely on another  microscopic  many-body approach, the Self-Consistent Green's Function method. 
This approach provides access to the single particle properties in a natural way. In particular, $n(k)$ can be obtained and, as a consequence, the kinetic energy can be studied. We will analyze 
how NN correlations affect differently the momentum distribution in symmetric and neutron matter and, therefore, which is the 
effect of the high momentum components on  the symmetry energy \cite{rios2012}.

In the SCGF method, a diagrammatic expansion is employed to solve for the in-medium one-body propagator, rather
 than for the energy of the system. For infinite matter, the method is conventionally applied at the ladder
 approximation level. SCGF calculations give direct access to microscopic properties related to the single-particle 
propagator. These include self-energies, spectral functions and momentum distributions, from which one can derive
 microscopic and bulk properties.  
The ladder approximation provides a microscopic description of short-range and tensor effects
 via a fully dressed propagation of nucleons in nuclear matter. This is achieved by: 
a) computing the scattering of particles via a $T$-matrix (or effective interaction) in the medium, 
b) extracting a self-energy out of the effective interaction and,
c) using the Dyson equation to build two-body propagators which are subsequently inserted in the scattering equation.
To solve this closed set of equations, a self-consistency procedure is required. 
The formalism is well established for two-body potentials and  its extension to include three-body forces has been only recently considered \cite{soma2008,arianna2013}. Here, all the SCGF results have been obtained with the AV18 NN interaction. 
Three-body forces are not included in this part of the work.

\begin{figure*}
\begin{center}
\resizebox{0.70\textwidth}{!}
{%
\includegraphics[clip=true]{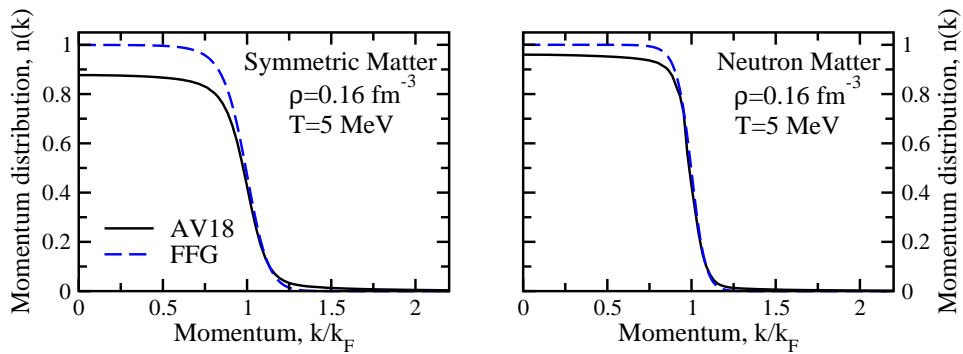}
}
\caption{(Color online) Momentum distribution of SNM (left panel) and 
PNM (right panel) obtained with the SCGF approximation for 
AV18 (full lines). The momentum distribution of the FFG (dashed line) 
obtained in the same conditions are also shown.} 
\label{fig:4}       
\end{center}
\end{figure*}

The bulk properties of nuclear matter and neutron matter are obtained within the SCGF approach
 through the Galitskii--Migdal--Koltun sum-rule:
\begin{equation}
\frac {E}{A} = \frac {\nu}{\rho} \int \frac {d^3k}{(2 \pi)^3} \int \frac {d \omega}{2\pi} 
\frac {1}{2} \left ( \frac {k^2}{2 m} + \omega \right ) {\cal A} (k,\omega) f(\omega) \, ,
\end{equation}
where $\nu=4(2)$ is the spin-isospin degeneracy of nuclear (neutron) matter, $\rho$ is the total density 
and $f(\omega) = [ 1 + e^{(\omega-\mu)/T} ]^{-1}$ is a Fermi--Dirac distribution. ${\cal A}(k,\omega)$ is 
the one-body spectral function which, loosely speaking, represents the probability of 
knocking out or adding a particle with a given single-particle momentum, $k$, and energy, 
$\omega$. The single-particle spectral function provides the full knowledge of the one-body propagator and 
gives access to the calculation of all the one-body properties of the system. For instance, 
the momentum distribution, $n(k)$, is obtained by convoluting the spectral function with a
 Fermi--Dirac factor: 
\begin{equation}
n(k) = \int \frac {d \omega}{2 \pi} {\cal A}(k,\omega) f(\omega) \, .
\end{equation}
The total density sum-rule,
\begin{equation}
\rho = \nu \int \frac {d^3k}{(2 \pi)^3} ~ n(k) \, ,
\label{eq:rho}
\end{equation}
is used to extract the chemical potential $\mu$. This can also be calculated using 
its thermodynamical definition, {\it i.e.,}  by a density derivative of the free-energy density. 
The agreement between these two  determinations of the chemical potential is taken 
as a test of the thermodynamical consistency of the approach. Notice that, to avoid pairing 
instabilities, the SCGF calculations have been performed at 
finite temperature (T=5 MeV) \cite{alm1996}. 

As mentioned earlier, correlations beyond the mean-field approximation have a  particularly clear
 manifestation in the momentum distribution \cite{rios2009}. A sizable depletion appears below the Fermi sea, while 
high-momentum components are populated. 
To illustrate this point, we show in Fig.\ \ref{fig:4}  the momentum distribution of SNM
(left panel) and PNM (right panel), at $\rho =0.16$ fm$^{-3}$ and $T=5$ MeV.
 The results obtained within the SCGF method for AV18 (solid lines) 
 are compared to the momentum distributions of the FFG in
the same conditions (dashed lines). 
The FFG is used here as a reference for the thermal effects, as deviations from the step
function give a measure of the importance of finite temperature.  A common characteristic 
 of the SCGF and the FFG $n(k)$  is the softening of the distribution around the Fermi surface, $k=k_F$, 
associated to the finite temperature.  Notice also that the density $\rho=0.16$ fm$^{-3}$ does not correspond 
to the saturation point of the AV18 potential. Actually, AV18 within SCGF approach
 at T=5 MeV, gives  a saturation density $\rho= 0.19 $fm$^{-3}$, smaller than the one obtained
in BHF with AV18 (see Table \ref{tab:3a}). 
  The inclusion of three-body forces should
 improve the saturation properties (as it happens in the BHF approach) without qualitative changes in the
 isovector properties \cite{soma2008}.

Correlation effects in the momentum distribution are substantially different in SNM  than in PNM
\cite{rios2009}. The effects of the tensor component in the 
$S-D$ channel in the isospin saturated system induce a large amount of correlations. Consequently,
 the Fermi surface is quite more depleted for SNM than for PNM 
(compare the left  and right  panels in Fig.\ \ref{fig:4}).  Characteristic values for these
 depletions are obtained from the occupation at zero momentum, namely, $n(0) \sim 0.87 $ for
 SNM and $n(0) \sim 0.96$ for PNM. As the momentum distribution is normalized to the
 total density (see Eq.\ (\ref{eq:rho})), 
the high-momentum components are also rather different for
 both systems at the same density. A useful way to characterize these differences is to look
at the integrated strength over  different regions of momenta:
\begin{equation}
\phi_m(k_i,k_f) = \frac {\nu}{2 \pi^2 \rho} \int _{k_i}^{k_f} dk k^m n(k) \, .
\end{equation}
The integral with $m=2$ represents the fractional contribution of a given momentum 
region to the total density, while the integral with $m=4$ is related to the total kinetic 
energy of the system. 

\begin{table}
\caption{Contributions of different momentum regions to the integrated strength 
with $m=2$ (columns 2 and 5), kinetic (3 and 6, in MeV) and total energies 
(4 and 7, in MeV). SNM (PNM) results are presented in columns 2, 3 and 4 (5, 6 and 7). 
Rows 3 to 6 show SCGF results with the AV18 NN interaction, 
whereas rows 7 to 10 correspond to the FFG results. 
All results are computed at $\rho=0.16$ fm$^{-3}$ and $T=5$ MeV.}
\label{tab:11}
\begin{center}
\begin{tabular}{l | lll | lll |}
\phantom{a} & \multicolumn{3}{ c }{SNM} & \multicolumn{3}{ | c |}{PNM} \\
\hline
  $(k_i, k_f)$   & $\phi_2$  & K/A & E/A & $\phi_2$  & K/A & E/A    \\
\hline\noalign{\smallskip}
\hline
    (0, $k_F$)  & 0.755 & 15.6 & -7.65 & 0.863 & 28.7 & 11.6   \\
    ($k_F$, $2k_F$)  & 0.194 & 11.4 & -1.00 & 0.119 & 10.3 & 3.24   \\
    ($2 k_F$, $\infty$)  & 0.051 & 14.5 & -1.29 & 0.018 & 7.16 & 0.32   \\
    (0, $\infty$)  & 1.00 & 41.5 & -9.94 & 1.00 & 46.2 & 15.2  \\
\hline\noalign{\smallskip}
\hline
    (0, $k_F$)   & 0.861 & 17.7 & 17.7 & 0.912 & 30.4 & 30.4  \\
    ($k_F$, $2 k_F$) & 0.139 & 6.00 & 6.00& 0.089 & 5.75 & 5.75  \\
    ($2 k_F$, $\infty$)  & 0.00 & 0.00 & 0.00 & 0.00 & 0.00 & 0.00 \\ 
    (0, $\infty$)  & 1.00 & 23.7 & 23.7 & 1.00 & 36.2 & 36.2  \\
\hline
\end{tabular}
\end{center}
\end{table}

In Table \ref{tab:11}, we report the integrated strengths with $m=2$ for  SNM (columns 2 to 4) and PNM (columns 5 to 7) at 
$\rho=0.16$ fm$^{-3}$ and $T=5$ for the AV18 
 potential (rows 3 to 6) and the FFG (rows 7 to 10). As expected, in SNM there is a substantial depletion of states
 below the Fermi surface, {\it i.e.,} only $\sim 75 \%$ of the strength is in the region between 0 and $k_F$. Part of the depletion
 has thermal origin, and the comparison with the FFG in the same momentum region suggests that between
 $1/2$ and $2/3$ of the integrated depletion comes from the softening of the Fermi surface due to finite
 temperature. The effect of correlations is also important in populating states beyond the Fermi surface: 
for SNM (PNM) there is still a 3-5 $\%$ (1-2$\%$) of strength in the region $k> 2 k_F$. 

The energy per particle is also affected by short-range and tensor correlations. In particular, the kinetic energy,
\begin{equation}
\frac {\langle K\rangle}{A} = \frac{\nu}{\rho} \int \frac {d^3k}{(2 \pi)^3}  \frac {\hbar^2 k^2}{2 m} n(k) \, ,
\end{equation}
noticeably  increases with  respect to the FFG due to the population of high-momentum components. 
The contributions of the different momentum regions to the kinetic energy of nuclear an neutron matter are 
also reported in columns 3 and 6 of Table \ref{tab:11}. 
The first thing to notice is that the total integrated  values (see rows 4 and 8 of Table \ref{tab:11}) of the correlated kinetic energies are larger 
than those of the FFG. For SNM,  dynamical correlations produce and increase of $17$ MeV with respect to the FFG,
 while the increment is only $10$ MeV for neutron matter, in agreement with the BHF estimations.
Overall, this reinforces the idea that correlations play a smaller role in PNM than in SNM.
 Paying attention to the different regions, we see that the momentum components beyond $k_F$ amount to
 more than $50 \, \%$ of the total in SNM, while in PNM they account for around $25 \, \%$ . 
In contrast, for the FFG at this temperature, the contributions of states above $k_F$ is less than $25 \, \%$ for SNM and less than
$15 \, \%$ for PNM.  The FFG strength above the Fermi surface is due to thermal effects, which for low temperatures 
are mainly localized within a small region around $k_F$. As a consequence, there is almost no 
contribution beyond $2k_F$. Therefore, this contribution in the interacting case can be entirely attributed to NN 
correlations. In the case of AV18, which is considered to be a hard interaction, the 
contribution beyond $2k_F$ is even larger than that between the Fermi surface and $2 k_F$. 
Similar analysis have been also performed for other realistic two-body potentials in Ref.~\cite{rios2012}.

\begin{figure}[t]
\begin{center}
\resizebox{0.40\textwidth}{!}
{%
\includegraphics[clip=true]{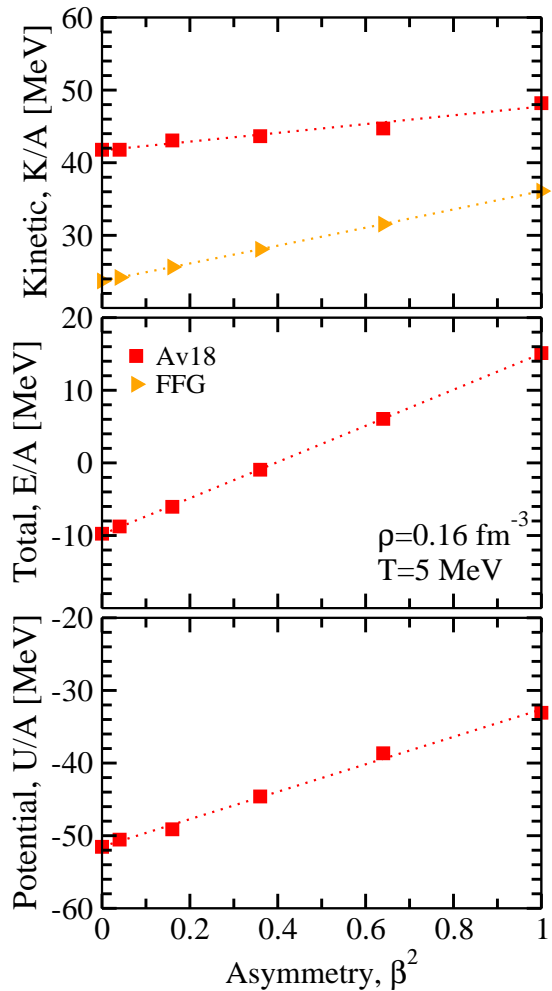}
}
\caption{(Color online) Dependence of the kinetic (upper panel), total (central panel)  and potential (lower 
panel) energy on the asymmetry 
parameter for the AV18 interaction (squares) at $\rho=0.16$ fm$^{-3}$ and $T=5$ MeV. The triangles
in the upper panel give the energy of the FFG in the same conditions. Dotted lines are linear regressions to guide the eye.}
\label{fig:5}       
\end{center}
\end{figure}

To compute the symmetry energy as the difference of the PNM and SNM energies, 
we have to resort  to the quadratic dependence of the total energy on the isospin asymmetry. 
This  quadratic behavior has been validated in Fig.\ \ref{fig:2} for the total energy calculated in BHF. 
The SCGF formalism has been generalized to isospin asymmetric systems and hence we can
directly check the quadratic asymmetry dependence of the energy \cite{frick2005}.  
Moreover, to explore the influence of the high momentum components
 on the symmetry energy, we have decomposed the symmetry energy in its kinetic and potential
 energy parts.
In the BHF approach, we have used the Hellmann--Feynman theorem to evaluate the kinetic energy. 
In the SCGF method, the kinetic energy is directly accessible from the momentum distribution
and the potential energy can be found from the Galitskii--Migdal--Koltun sum-rule. 
Consequently, and as a first step before computing isovector properties, we validate the quadratic behavior of both
pieces of the total energy with the asymmetry parameter. 
 
In Fig.~\ref{fig:5}, we show
the kinetic, potential and total energy for the AV18 potential at $\rho=0.16$ fm$^{-3}$ an 
$T=5$ MeV as a function of $\beta^2$. In general, the three components seem to have a well-defined parabolic 
dependence on the asymmetry parameter. The slope of the linear regression reduces to the different components 
(kinetic, potential and total) of the symmetry energy when the parabolic approximation holds exactly. In the
 upper panel of Fig.~\ref{fig:5} we compare the kinetic energy provided by the AV18 potential and the one of the FFG
 (triangles). As expected, the correlated kinetic energy is larger than the FFG at all asymmetries.
However, the isospin dependence  is different than the one of the FFG. While the kinetic energy of AV18 in SNM 
is $K/A \sim 42$ MeV and that of PNM is $\sim 46$ MeV, the FFG gas provides $\sim 24$ MeV  and
$\sim  36$ MeV respectively. The 
difference between the correlated kinetic energies associated to PNM and SNM is smaller than for the FFG. 
This agrees with the estimate provided by the BHF approach at $T=0$. 
The small value of the kinetic symmetry energy (clearly smaller than the FFG estimation) is a very
 noticeable aspect 
of the decomposition of the symmetry energy in a kinetic and a potential energy component.
 The origin of this behavior can be related to the tensor and short range repulsive components of 
the NN force,
 that when acting on SNM, induce large correlations and produce an important
 renormalization of the kinetic energy with respect to the FFG. The absence of this
 components in PNM (some partial waves are suppressed due to the Pauli principle) reduces 
the relative enhancement of the kinetic energy. Consequently, the difference in total
 kinetic energies of PNM and SNM is smaller for the correlated case
 than for the FFG value.  
 
\begin{figure}
\begin{center}
\resizebox{0.40\textwidth}{!}
{
\includegraphics[clip=true]{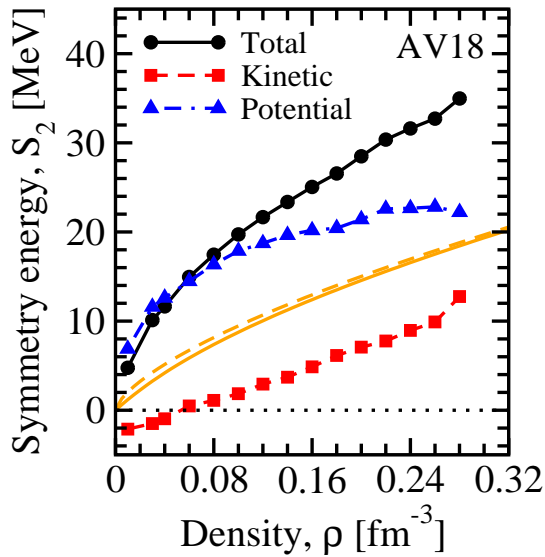}
}
\caption{(Color online) Components of the symmetry energy for the AV18 interaction calculated in the SCGF
approach, as a function of density at $T=5$ MeV. Circles, squares and triangles represent the 
total, kinetic and potential contributions, respectively. The continuous (dashed) lines correspond
to the FFG symmetry energy at $T=5$ ($T=0$) MeV.} 
\label{fig:6}       
\end{center}
\end{figure}

The asymmetry dependence of the  total energy per particle is driven by a balance
 between the kinetic and potential terms. The size of both contributions is density
 dependent but, at $\rho = 0.16$ fm$^{-3}$, the potential  term dominates the 
isospin dependence, as seen in Fig.~\ref{fig:6}. In other words, the potential energy contribution to
 the symmetry energy is $20.2$ MeV, while the kinetic energy part is only $4.9$ MeV. The total 
 value for the symmetry energy is $S_{2}= 25.1$ MeV, somewhat below the BHF result 
 including 3BFs. However, one should take into account 
that the BHF calculation is done at $T=0$ and computed at a saturation density of $0.187$ fm$^{-3}$. 
Regarding thermal effects, the symmetry energy of the FFG can provide an indication of their importance.
At $\rho=0.16$ fm$^{-3}$, the symmetry energy increases from $12.4$ MeV at zero temperature to 
$13$ MeV at $T=5$ MeV. Overall, this indicates a very small effect of temperature on the symmetry
energy. This is a result of the cancellation of the somewhat similar temperature dependences of 
symmetric and neutron matter \cite{rios2008}. 

The total symmetry energy predicted by the SCGF approach 
is a little below the currently accepted value of $\sim 32 $ MeV \cite{tsang2009}.
 In principle, the inclusion of 3BFs should bring the SCGF results closer to experiment. 
 A first estimation of the effect of 3BF can be obtained from the BHF calculations including 3BF. Around
$\rho = 0.16$ fm$^{-3}$, 3BF tend to increase the symmetry energy by $3-4$ MeV. A similar increase
in the SCGF case would bring the value closer to experiment. 

In general, we see that 
the symmetry energies provided by SCGF calculations tend to be smaller than the BHF ones with the same  
2NF  force. The origin of this difference  can be argued as follows. Compared to the BHF approach, the ladder summation
in the SCGF approach includes hole-hole diagrams as well as the full dressing of the 
intermediate propagators. It is known that both things have an overall repulsive effect in the total energy of the system
 with respect to the BHF values 
\cite{muether2000}. This repulsive effect is mostly associated to correlations and, since these are more relevant in SNM,
we expect more repulsion in SNM than in PNM. The difference in energies
between PNM and SNM is therefore reduced with respect to BHF and the SCGF symmetry energy becomes smaller than the BHF one. Note that, since this repulsive effect increases with density, the slope of the symmetry energy as a function of density is also expected to decrease. 

The density dependence of the kinetic and potential energy components of the symmetry energy 
is shown in Fig. \ref{fig:6}. The symmetry energy and its components grow steadily 
in the density range considered. The potential part is always larger in absolute value  than the kinetic one, and dominates the 
contribution to $S_2$. It is interesting 
to note that the kinetic symmetry energy becomes negative at densities around 0.04-0.08 fm$^{-3}$. 
One might expect thermal effects to be important in this regime, but the comparison with the FFG once again
demonstrates that finite temperature has a negligible effect on the symmetry energy. As a matter of fact, the comparison
between the symmetry energy of the FFG at $T=0$ (dashed line) and at $T=5$  MeV (solid line) shows that the differences 
are extremely small (less than $1$ MeV) in the whole density regime. 
As mentioned above, the small influence of the temperature on the 
symmetry energy is caused by the relatively similar thermal corrections of SNM and
 PNM \cite{rios2008}. When taking the difference of both energies, one eliminates practically the 
temperature dependence. Consequently, the negative kinetic symmetry energies at low densities 
can be considered a  NN correlation effect. This is similar to the BHF case around saturation density and
has also been confirmed in other many-body calculations. 

\section{ Summary and Conclusions}
\label{sec:5}
We have studied the density dependence of the symmetry energy within  the microscopic
 Brueckner--Hartree--Fock and the Self-Consistent Green's Function approaches, using
 the realistic AV18 NN potential as a starting point. In the BHF case, we have supplemented
 our calculations with the Urbana IX three-body force. The 
BHF calculations provide a symmetry 
energy, $E_{sym} = 34.3$ MeV, and a value of the slope parameter, $L = 66.5 $ MeV, compatible 
with recent experimental constraints. 
Using the Hellmann--Feynman  theorem, we have evaluated the separate contributions  of the different 
components of the NN interaction to the nuclear symmetry energy and to the slope parameter. 
This allows for a decomposition of the symmetry energy in a kinetic and potential  energy parts. The results
show that the potential  part gives the main contribution to both $E_{sym}$ and $L$ and that 
the kinetic energy contribution is very small. 
We have also performed a partial-wave as well as a spin-isospin channel decomposition of the potential
 part
of $E_{sym}$ and $L$, showing that the largest contribution is given by the spin-triplet
($S=1$) and isospin singlet  ($T=0$) channels.  All results  point  to the dominant role of the tensor
component of the NN force, which gives the largest contribution to both $E_{sym}$ and $L$. 

We have completed our analysis by an explicit calculation of the momentum distributions within 
the Self-Consistent Green's function approach. We have shown how  correlations affect
 differently the momentum  distributions of SNM and PNM.
 We have performed the analysis by quantifying the contribution  of momenta beyond the Fermi
 surface to the kinetic and total energies by using the Galitskii--Migdal--Koltun sum-rule.
 The change in the high momentum components  as the isospin asymmetry is modified 
 confirms the decrease of the kinetic energy component of the 
symmetry energy with respect to the free Fermi gas. Both changes, namely the decrease in the  kinetic energy component of the symmetry energy as well as the change in  $n(k)$ with respect 
to the uncorrelated step function, are used as an indicator 
of the presence of correlations in SNM and PNM. Also, both measures point towards the same
conclusion, namely that  
correlations play a smaller role in PNM than in SNM. Or, in other words, pure neutron matter is a less correlated system than symmetric nuclear matter.    
  
\section*{Acknowledgments}

This work has been supported by the Spanish MICINN Grant No. FIS2011-24154; 
the Generalitat de Catalunya Grant No. 2009SGR-1289; the Consolider Ingenio 2010 Programme 
CPAN CSD2007-0042; STFC, through Grants ST/I005528/1 and  ST/J000051/1;
the R\&DT projects \\ PTDC/FIS/113292/2009 and CERN/FP/123608/2011, developed under 
the scope of a QREN initiative; UE/FEDER financing, through the COMPETE programme;
and by the NEW COMPSTAR, a COST initiative.


\end{document}